\documentclass[letterpaper,aps,prb,floatfix,twocolumn,showpacs,superscriptaddress,10pt]{revtex4-1}
\usepackage{graphicx,times}
\usepackage{amssymb,amsmath}
\usepackage{multirow}
\usepackage{color}
\usepackage{epstopdf}
\usepackage{bm}% bold math
\usepackage{longtable}
\usepackage{lipsum}
\usepackage{braket}
\usepackage[colorlinks=true, urlcolor=blue, linkcolor=red, citecolor=blue, pdftex]{hyperref}
\hypersetup{pdfstartview={FitV}, pdfpagemode = UseNone, pdfpagelayout=SinglePage, bookmarksnumbered={true}}
\def\be{\begin{equation}}
\def\ee{\end{equation}}
\def\ba{\begin{eqnarray}}
\def\ea{\end{eqnarray}}

\usepackage[section]{placeins}

\begin{document}
% Last edited by Tapio 9.3.2018

\title{Excitation Energy Transport with Noise and Disorder in a Model of the Selectivity Filter of an Ion Channel}

\author{Amir Jalalinejad}
\affiliation{Dipartimento di Scienze Molecolari e Nanosistemi, Universita' Ca'Foscari Venezia, 
Edificio Alfa Campus Scientifico, via Torino 155, Venezia-Mestre I-3010, Italy}

\author{Hassan Bassereh}
\affiliation{Department of Physics, Isfahan University of Technology, Isfahan 84156-83111, Iran}

\author{Vahid Salari}
\affiliation{Department of Physics, Isfahan University of Technology, Isfahan 84156-83111, Iran}

\author{Tapio Ala-Nissila}
\affiliation{Department of Applied Physics and COMP CoE, Aalto University School of Science, P.O. Box 11000, 00076 Aalto,
Espoo, Finland}
\affiliation{Interdisciplinary Centre for Mathematical Modelling and Departments of Mathematical Sciences and Physics, 
Loughborough University, Loughborough, Leicestershire LE11 3TU, UK}

\author{Achille Giacometti}
\affiliation{Dipartimento di Scienze Molecolari e Nanosistemi, Universita' Ca'Foscari Venezia, Edificio Alfa Campus Scientifico, via Torino 155, Venezia-Mestre I-3010, Italy}

\date{March 9, 2018}
%%%%%%%%%%%%% Abstract %%%%%%%%%%%%
\begin{abstract}
Selectivity filter is a gate in ion channels which are responsible for the selection and fast conduction of particular ions across the membrane 
(with high throughput rates of $10^8$ ions/sec and a high 1:$10^4$ discrimination rate between ions). It is made of four 
strands as the backbone, and each strand is composed of sequences of five amino acids connected by peptide units H-N-C=O in which the main molecules in the 
backbone that interact with ions in the filter are carbonyl (C=O) groups that mimic the transient interactions of ion with binding sites during ion conduction. It has been 
suggested that quantum coherence and possible
emergence of resonances in the
backbone carbonyl groups may play a role in mediating ion
conduction and selectivity in the filter. Here, we investigate the influence of noise and disorder on the efficiency of excitation energy transfer (EET) in a linear harmonic chain of $N=5$ sites with 
dipole-dipole couplings as a simple model for one P-loop strand of the selectivity filter backbone in biological ion channels. 
We include noise and disorder inherent in real biological systems by including spatial disorder in the chain, and random noise within a
weak coupling quantum master equation approach. Our results show that disorder in the backbone considerably reduces EET, but the addition of
noise helps to recover high EET for a wide range of system parameters. Our analysis may help for better understanding of the coordination of ions in the filter as well as the fast and efficient functioning of the selectivity filters in ion channels.
\end{abstract}

\pacs{pacs}

\maketitle
%%%%%%%%%%%%%  Introduction %%%%%%%%%%%%%%%%%%%
\section{Introduction}

{\it Excitation Energy Transfer}  (EET) has been recently investigated as one of the most important problems in physical and biological systems, e.g. in photosynthesis~ \cite{felming2007,scholes2010,fleming.Science2007,engel.ProcNatlAcadSciUSA2010,hulst.Science2013,CambridgeUniversityPressCambridge2013}, 
charge transfer in DNA \cite{wessely.Nature2001} or motional excitation in $\alpha$-helix structures.\cite{Cruzeiro.J.Chem.Phys.2005, may.Chem.Phys.2007}
Noise and disorder are inherent factors in EET in biological and complex systems and they originate from various types of internal and external 
sources and affect the system dynamics. Recent studies have indicated that
disorder and noise can actually {\it enhance} EET in some cases. This is in contrast to the standard scenario in quantum mechanical systems where strong enough disorder leads
to localization of the wave functions \cite{anderson. Phys. Rev.1958} and thus vanishing of the related transport coefficient. 
In particular, it has been shown that disorder increases EET in a symmetric network because of invariant disconnected subspaces~\cite{plenio.The Journal of Chemical Physics2009}, 
or complex networks suffering from  localization~\cite{anderson. Phys. Rev.1958,hassan2}. 
Also, disorder can optimize transport efficiency of a chromosphere inside a sphere 
\cite{rabitz.The Journal of Chemical Physics2013,buchleitner.Acta Physica Polonica A 2011, buchleitner.Phys. Rev. E 2011}. 
Energy transport efficiency enhancement has also been observed due to disorder on site energies in Ref. \onlinecite{yasser Omar.arXiv:1312.6989}.\\

%%%%%%%%%%%%%%%%%%% FIG 1 %%%%%%%%%%%%%%%%%%%%%%%
\begin{figure}[h]
   \includegraphics[width=7cm]{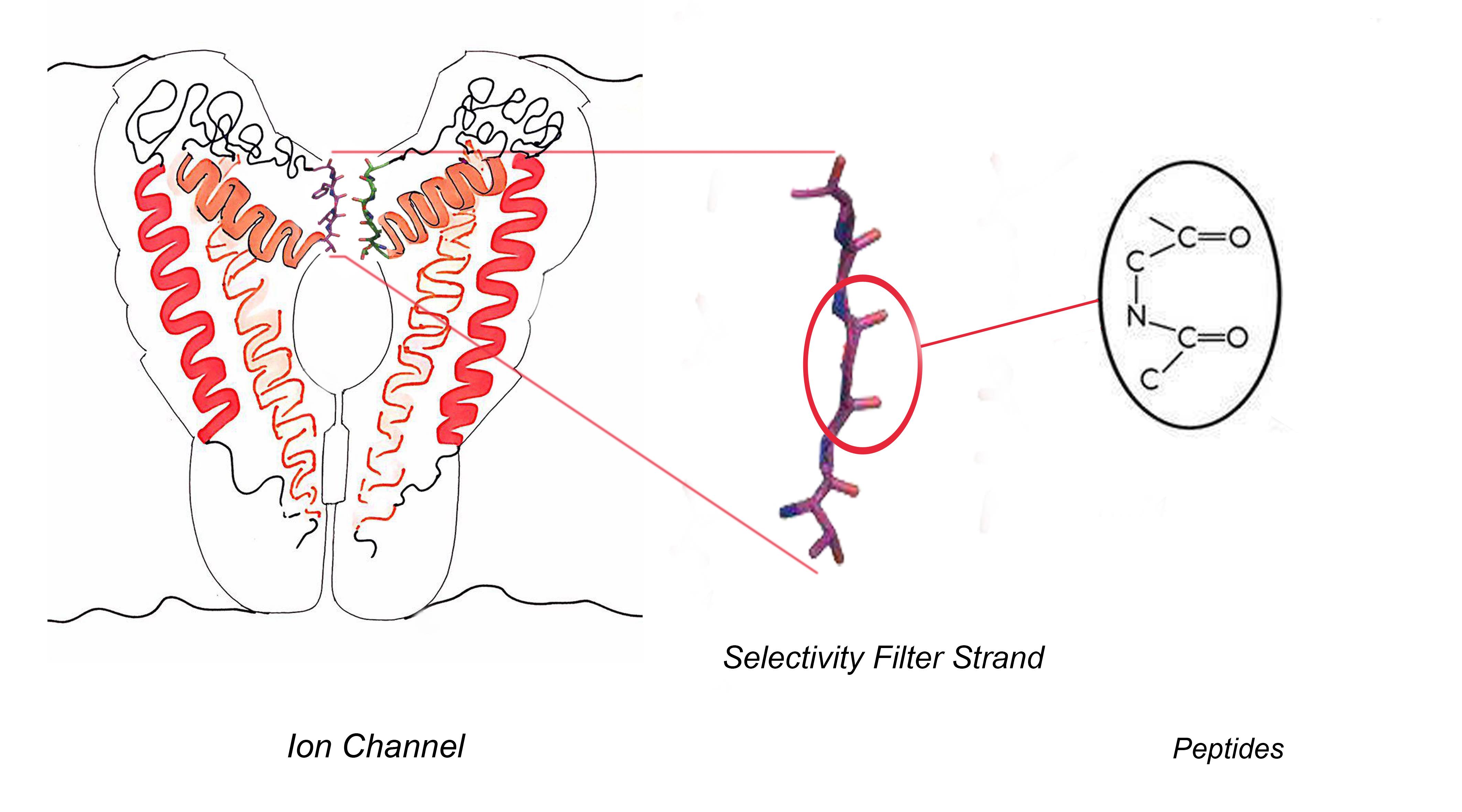}
       \caption{ (Colour online) Schematic of a backbone strand of a selectivity filter. In the structure of the 
       selectivity filter N-C=O is an amide group and C=O a carbonyl group. Each C=O pair 
       can be considered as an electric dipole that interacts with the C=O bonds in the neighboring sites.}
            \label{fig:fig1}
\end{figure}
%%%%%%%%%%%%%%%%%%%%%%%%%%%%%%%%%%%%%%%%%%%%%%%%

It has been shown that quantum effects in ion channels may play role in ion transfer as well as the efficient functioning of the ion channel 
\cite{plenio.New Journal of Physics2010, Bernroider1, Bernroider2, Semiao, Vahid1, Vahid2, Vahid3, iiiii, hassan}. Two issues are of particular interest within this framework. 
The first issue is related to quantum effects in ion transport \cite{plenio.New Journal of Physics2010}. 
The second issue is related to quantum effects in the selectivity filter backbone, and will be the main topic of the present study.

The selectivity filter is a gate in ion channels which are responsible for the selection and fast conduction of particular ions across the membrane with very high throughput rates 
($10^8$ ions /sec) and high (1:$10^4$) discrimination rate between ions. It is made of four strands as the backbone, where each strand is composed of a sequence of five amino acids 
connected by peptide units H-N-C=O (see Fig.~\ref{fig:fig1}) \cite{iiiii}. The main molecules in the backbone that interact with the ions in the filter are carbonyl (C=O) groups, which mimic the transient interactions of ion with binding sites of the selectivity filter during ion conduction \cite{Vaz2}.

It has already been remarked that the selectivity filter backbone is not stiff, as expected in classical deterministic models,  
and therefore vibrational excitations in K ion channels in which quantum coherence and possible emergence of resonances at the 
picosecond (ps) time scale in the backbone carbonyl groups may play a role in mediating ion conduction and selectivity in the selectivity filter \cite{Vaz2}. 
It has been shown \cite{plenio.New Journal of Physics2010, Vaz2} that the Hamiltonian describing the dynamics of a chain of molecules that is subjected to an axial 
Coulomb potential in the selectivity filter can be written as a set of coupled quantum harmonic oscillators connected to a source and a sink where excitations enter 
and leave the system \cite{plenio.New Journal of Physics2010}. The main molecules in the backbone that interact with ions in the filter are the carbonyl (C=O) groups.
Here, we use this modeling scheme to explore the influence of noise and disorder on EET.
We note that the detailed operational principles of the selectivity filter are still unknown due to its small size, small values of the energy exchanges, and fast ps time scales. 
This makes it very difficult to set up a realistic model especially in view of lack of experimental data on EET in the backbone 
as well as probing the dynamics on the relevant time scales. 
To this end we use a simplified model of a single strand in the backbone of the selectivity filter (and not the whole ion channel) to 
get better understanding of the functioning of the system from a quantum mechanical point of view.
 
The backbone of a selectivity filter is composed of four similar (linear) strands that regulate the ion transport via the interaction of carbonyl groups in the backbone and ions inside the filter. 
The important question is: how does a flexible and possibly spatially disordered structure such as the filter achieve ion selectivity and high throughput at the same time?
Within the simplified picture described above, we consider here only one of the strands (in the absence of ions) modeled as a linear chain of five sites, 
where each site plays the role of a peptide unit linked to an amino acid, and each site is a two-state system (with a ground state $|g\rangle$ and an excited state $|e\rangle$) 
which can be excited, e.g. due to the presence of an ion. The interaction between any two sites is described to be the inverse power law dipole-dipole interaction \cite{ali}.
As in any biological system, the selectivity filter backbone is coupled with the external environment that will be mimicked here by including a noise term in the Hamiltonian. 
In doing this, we will extend previous work \cite{hassan} in that the couplings between the sites are changing in time due to the intrinsic vibrations in the system.  

The outline of the rest of the paper is then as follows. Section \ref{model} describes the model, while Section \ref{results} includes the numerical results of our analysis. 
In Section \ref{discussion} we try to make contact with real systems and discuss the limitations of the model. 
Finally, Section \ref{conclusions} will close the paper with some conclusions and future perspectives. 

%%%%%%%%% Section {Our Model: Structure and Transport }%%%%%%%%%%%%%%%%%

\section{Model for Structure and Transport}
\label{model}
We consider a linear harmonic chain consisting of $N$ equivalent two-level sites. The energy injected into such a system, usually into the first site, 
can be transported along the chain through the interaction between the neighbouring sites. This energy transport can be modelled by the following 
general tight-binding Hamiltonian in the single excitation sector of the Hilbert space~\citep{plenio.New.J.Phys.2008,ali}:
\begin {equation}\label{tb1} 
H=\sum_{n=1}^{N} \varepsilon_{n}|n\rangle\langle n|+ \sum_{n=1}^{N-1}J_n(t)(|n\rangle\langle n+1|+|n+1\rangle\langle n|),
\end {equation}
where $\vert n \rangle$ denotes an excited atom on the $n^{\rm th}$ site (e.g., for a three-site system we have 
$\vert 1 \rangle=\vert e \rangle \otimes \vert g \rangle \otimes \vert g \rangle$ and $\vert 2 \rangle=\vert g \rangle \otimes \vert e \rangle \otimes \vert g \rangle$), 
$\varepsilon_{n}$ is the excitation energy at site $n$, and $J_n(t)$ is the hopping integral generated by the coupling between the neighbouring atoms, whose details will be explained below. 
Both $\varepsilon_{n}$ and $J_n(t)$ have the units of energy. Following Ref.\cite{ali} variables will be rescaled to have $\varepsilon_{n}=1$  as well as $\hbar=1$ in Eq. (\ref{tb1}).

We consider the chain as an open quantum system \citep{Huelga} and model the dynamics using the standard Markovian master 
equation approximation in the weak coupling limit, 
with Lindblad superoperators to account for the energy loss due to dissipation~\citep{ali}:
\begin {equation}\label{loss} 
L_{\rm diss}\rho=
\gamma_{\rm diss} \sum_{n=1}^{N} [2{\sigma}^-_n\rho{{\sigma}^+_n}-\big\{{{\sigma}^+_n{\sigma}^-_n,\rho}\big\}],
\end {equation}
in which $\rho$ is the total density matrix, ${\sigma}^+({\sigma}^-)$ denotes the creation (annihilation) operator of the 
excitation at site $n$, $\gamma_{\rm diss} = 0.001$ is the rate of dissipation (the reason for this value will be explained below). 
In Eq. (\ref{loss}) we have used the definition  $\big\{{A,B}\big\}=AB+BA$.\cite{footnote1}

To calculate the fraction of the initial excitation energy transferred along the chain (and not lost because of dissipation), 
we add an additional {\emph {sink}} to the end of the chain. The sink is populated via an irreversible interaction with the end site $N$. 
Such an irreversible process can be formulated in terms of the Lindblad operators as \citep{hassan}:
 \begin {equation}\label{sink} 
 L_{\rm sink}\rho=\gamma_{\rm sink}[2{\sigma}^+_{N+1}{\sigma}^-_N\rho{\sigma}^+_N{\sigma}^-_{N+1} 
 -\big\{{{{\sigma}^+_N{\sigma}^-_{N+1}\sigma}^+_{N+1}{\sigma}^-_N,\rho}\big\}],
\end {equation}
where $\gamma_{\rm sink}$ denotes the absorption rate of the sink that will be set to $\gamma_{\rm sink}=0.1$ in most of our computations.

The dynamical process of energy transfer starts at $t=0$ by injection of an excitation to the first site $N=1$, {i.e.}  
$\rho(0)=\vert 1\rangle \langle 1 \vert$. We are interested in the amount of excitation that has been transferred to the sink after time $t$. 
To this end, we calculate the \emph{sink population} given by \citep{plenio.New Journal of Physics2010}: 
\begin{equation}\label{pop}
\eta(t)=2\gamma_{\rm sink} \int^t_0 \langle N \vert \rho(t) \vert N \rangle dt.
\end{equation}
The total density matrix $\rho(t)$ can be obtained as the solution of the following master equation:
\begin{equation} 
\frac{\partial\rho}{\partial{t}}=i[H,\rho]+L_{\rm diss}\rho+L_{\rm sink}\rho.
 \label{lin}
 \end{equation}
We focus here in the steady state and therefore calculate the sink population in the long time limit  $\eta_{\infty} = \eta(t \to \infty)$ 
and call this quantity the \emph{system efficiency} \citep{lloyd.Physical Review E2012}. In fact, due to dissipation some part of the energy is dissipated at each site through energy transfer during time, and therefore the whole energy cannot be received in the sink after the saturation time. 

We consider a system in which the excitation energy hop from one site to the next one occurs through the dipole-dipole interaction. 
For the hopping integrals we then have~\citep{ali}:
\begin{equation}\label{dd}
J(t)= J_0/{d_n}^3.
\end{equation}
In Eq. (\ref{dd}) $J_0$ is determined by the magnitude the dipole moments (here assumed to be constant) 
and physical constants. Due to thermal motion the chain should
be allowed to fluctuate which leads into time dependence of the dipole-dipole interaction.\cite{footnote2} 

\begin{figure}
\includegraphics[width=7cm,height=7cm]{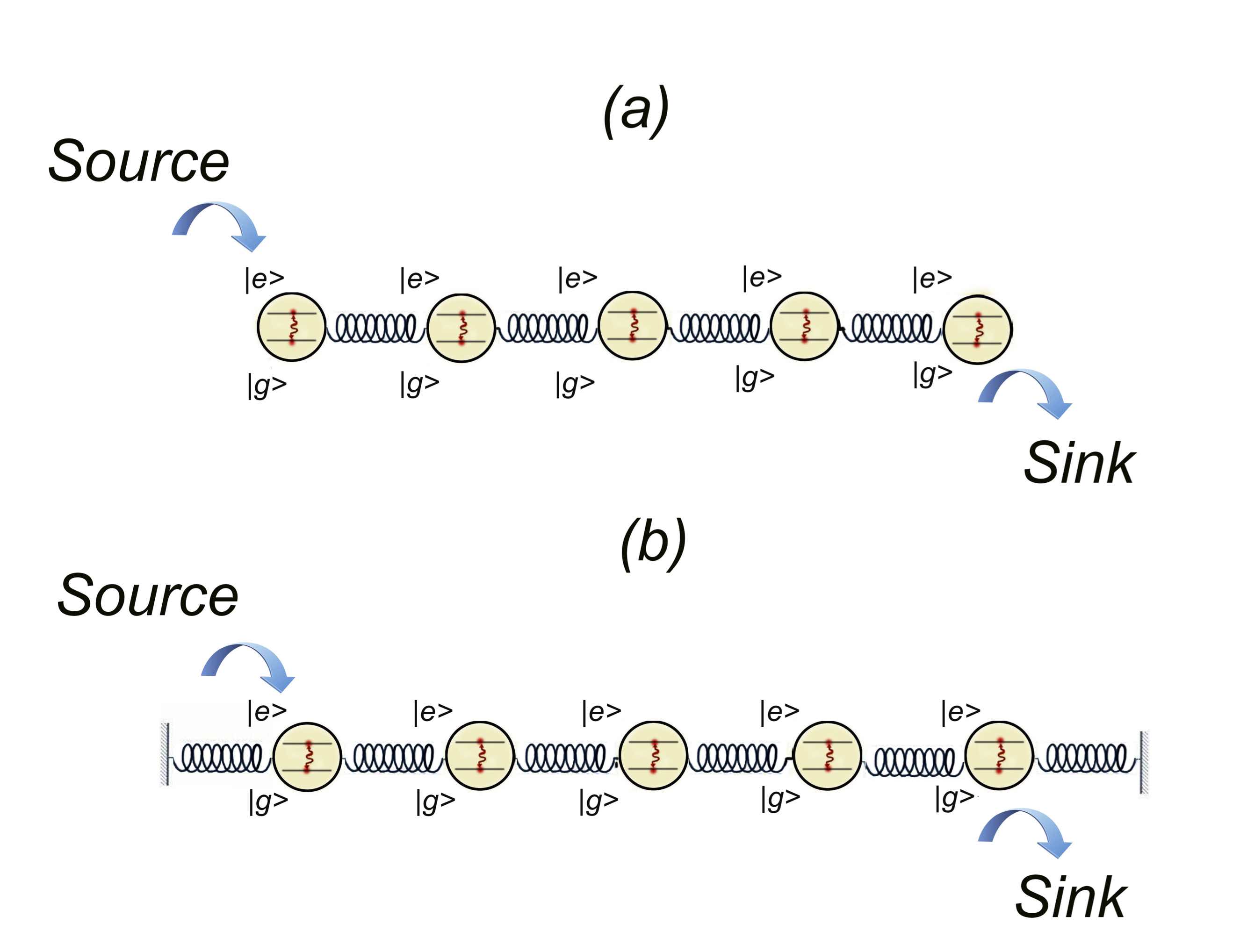} 
 \caption{(Colour online) Schematic representation of a $N=5$ harmonic chain of two-level sites with (a) the breathing model where the ends are free, 
 and (b) the confined model where the end atoms are confined. Here $\ket{g}$ and $\ket{e}$ stand for the ground and excited states, respectively.}
\label{fig:fig2}
\end{figure}

We assume that the time dependence of the position of each site is given by $x_n = a_n \cos(\omega t)$ and thus 
the distance between neighbouring sites follows the simple equation 
$d_n={d_0 }[1 - 2a_n \sin({\omega}t+\phi)]$, where $d_0$ is the equilibrium distance, and $a_n$ the amplitude of the longitudinal oscillations
with frequency $\omega$. ${a_n}$  is the ratio of the amplitude of longitudinal oscillations of each atom to  $d_0$ and $\omega$ is the frequency of these oscillations. 
To obtain $a_n$ and $\omega$ for a given chain of mass and springs we have to calculate the normal modes. 
For a finite chain the end atoms can be considered to be free which we call the \emph{breathing model}, or confined in space which we
call the \emph{confined model}. Figure \ref{fig:fig2} shows the breathing and confined models for a chain of length $N=5$. 
In the selectivity filter, each peptide unit H-N-C=O is connected to an amino acid and thus the confined model is the relevant case here.
In our numerical simulations all energies, time scales, and rates will be expressed in units of $J_0 = 1$.

%%%%%%%%%%%%%%%%%%%%% Figure 2 %%%%%%%%%%%%%%%%%%%%%%%%%%%%%%%%%%%%%

%%%%%%%%%%%%%%%%%%%%%%%%%%%%%%%%%%%%%%%%%%%%%%%%%%%%%%%%%%%%%%%%%%%%%%
\section{Results}
\label{results}
%%%%%%%%%%%%%%%%\Section {EET In a chain of length N=5} %%%%%%%%%%%%%%%%

\subsection{EET in a chain of length $N=5$}

We first consider a system of five masses and six springs between two walls, as shown in Fig. \ref{fig:fig2}(b), where the masses are all equal to 
$m=1$ and the spring constants are all equal to $k=1$. Let $ x_1, x_2, x_3, x_4, x_5$ be the displacements of the masses relative to their equilibrium positions. 
Thus we can write the equation of motion for each site as
\begin {equation}\label{eom}
m\ddot{x}_{n}+k(x_{n}-x_{n-1} )+k(x_{n}-x_{n+1} )=0,
\end {equation}
which immediately gives us the normal modes shown in Table \ref{tab11}. In obtaining these values we have one degree of freedom 
which we take it to be the amplitude of fifth site $a_{5}=0.25$. These normal modes are used to calculate the time-dependent hopping parameters 
$J_n(t)={J_0}/{d^3_n(t)}$, where $d_n=d_{0}-(x_n-x_{n+1})$, $J_0=1$, and $\:x_n=a_n\sin(\omega t)$ for $n=1,2,3,4,5$.\cite{python}

\begin {table}
\centering

\begin {tabular}{| c | c | c | c | c | c | c |}
\hline
$\omega$ & $a_1$ & $a_2$ & $a_3$& $a_4$& $a_5$ 
\\ \hline
$\omega_1=\sqrt{2-\sqrt{3}}$         & 0.25   & 0.25$\sqrt{3}$     & 0.5   &0.25$\sqrt{3}$           &0.25  
\\ \hline
 $\omega_2=1$                             & -0.25  &-0.25               &0.0    &0.25         &0.25 
\\ \hline
$\omega_3=\sqrt{2}$                     & 0.25   &0.0                 &-0.25  &0.0        &0.25  
\\ \hline
$\omega_4=\sqrt{3}$                     &-0.25  &0.25                &0.0    &-0.25       &0.25 
\\ \hline
$\omega_5=\sqrt{2+\sqrt{3}}$       & 0.25   &-0.25$\sqrt{3}$     &0.5    &-0.25$\sqrt{3}$          &0.25  
\\ \hline
    
    \end {tabular}
    \caption{Normal modes of a $N=5$ confined harmonic chain with $\omega_{0}=1$ and $a_5=0.25$.}
%\label{tab11}
\end {table}
%%%%%%%%%%%%%\section{EET in Ordered and Disordered Chains} %%%%%%%%%%%%%%

\subsection{EET in Ordered and Disordered Chains}
\label{static}

In this section we study the quantum excitation energy transfer in ordered and disordered harmonic chains in which the distances 
between the sites are time-dependent according to the model presented in the previous section. The {\it ordered harmonic chain} is one where the equilibrium distances between
 all sites are equal. In a {\it disordered chain} the equilibrium distances can vary.
 
The coupling between two interacting sites is expressed by Eq. (\ref{dd})  
where $d_0=1.21$ for the ordered chain ~\cite{yyyy}. For the disordered chain we use 
the relative ratios of distances based on the real data for amino acids in one strand of the backbone of the selectivity filter in KcsA ion channels \cite{hassan}. 
To this end, we denote the $n$ dependent nearest neighbour equilibrium distances by $d_{0: n, n+1}$. 
In this structure, the distances between neighbouring sites are $r_{0:1,2}=2.6 {\buildrel _{\circ} \over {\mathrm{A}}}, 
r_{0:2,3}=3.1 {\buildrel _{\circ} \over {\mathrm{A}}}, r_{0:3,4}=2.6 {\buildrel _{\circ} \over {\mathrm{A}}}$, and $r_{0:4,5}=4.3 {\buildrel _{\circ} \over {\mathrm{A}}} $ ~\cite{10}. 
Normalizing these distances with the minimum we obtain $d_{0:1,2}=1,d_{0:2,3}=1.19, d_{0:3,4}=1$ and $d_{0:4,5}=1.65$.

\begin{figure}
\includegraphics[width=8cm,height=7cm]{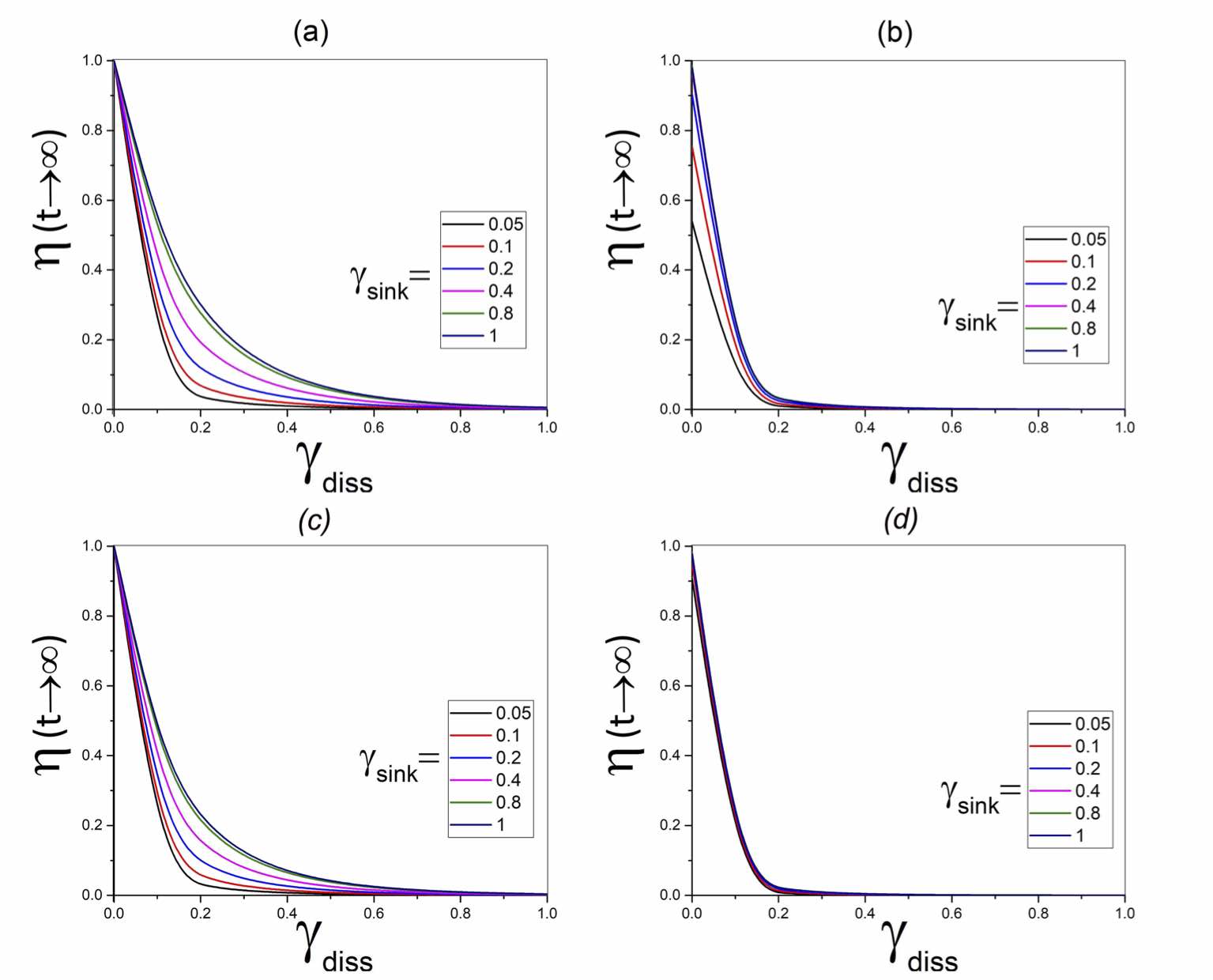} 
 \caption{(Colour online) The asymptotic sink population $\rho_{\rm sink}= \eta(t \to \infty)$ as a function of the 
 dissipation coefficient $\gamma_{\rm diss}$, for different values of $\gamma_{\rm sink}$, (a) order without noise (b) disorder without noise (c) order with noise ($\gamma_{\rm deph}=0.2$), and (d) disorder with noise ($\gamma_{\rm deph}=0.2$). The above diagrams are plotted for the normal frequency $\omega_1$.}
\label{fig:fig3}
\end{figure}

%\subsection{Dissipation}
The dissipation effect is not our primary concern in the present work that will mainly address the behavior of the system under the effect of disorder and noise. 
However, in order to select a reasonable value for the dissipation coefficient $\gamma_{\rm diss}$, we have first considered an ordered chain with no noise, 
and computed the long time sink population $\gamma(t \to \infty)$
as a function of $\gamma_{\rm diss}$ (cf. Eq. (\ref{loss})), and for different values of 
$\gamma_{\rm sink}$ (cf. Eq.(\ref{pop}). The results are shown in Figure \ref{fig:fig3}, where the effect of different values of the dissipation factor on 
the sink population in the asymptotic state is reported. We note that when the dissipation is high ($\gamma_{\rm diss} \approx 1$), the 
EET behavior is significantly reduced. This suggests to use a small value ($\gamma_{\rm diss}=0.001$)
in all the following calculations in order to unravel the effect of noise and disorder with a finite dissipation rate.

%%%%%%%%%%%%%%%%%%%%% Figure 3 %%%%%%%%%%%%%%%%%%%%%%%%%%%%%%%%%%%%%%%%
%%%%%%%%%%%%%%%%%%%%%%%%%%%%%%%%%%%%%%%%%%%%%%%%%%%%%%%%%%%%%%%%%%%%%%

In Fig. \ref{fig:fig4} (and Fig.1S in supplementary file)  we show our numerical results for the time dependence of the site and sink populations for different normal modes and related amplitudes corresponding to Table 1.  Only results for the first ($\omega_1$) and the last ($\omega_5$) are here reported for simplicity. Internal modes ($\omega_2$, $\omega_3$, and $\omega_4$) can be shown to provide consistent results. We can see that for the ordered system in Fig. \ref{fig:fig4} (and Fig.1S) (a), there is little dependence of the asymptotic sink population on
the modes. In particular, for the three lowest modes the steady state populations are almost identical, and decrease slightly for the higher modes. 
By contrast, for the disordered model of Fig.  \ref{fig:fig4} (and Fig.1S)(b), the overall sink populations are much smaller. This can be rationalized in terms of the localization of the 
wave functions due to spatial disorder, hence significantly affecting the EET efficiency\cite{hassan,footnote3}.
%%%%%%%%%%%%%%%%%%%%%%%%%%%%  Figure 4 %%%%%%%%%%%%%%%%%%%%%%%%%%
\begin{figure*}
   \includegraphics[width=16cm,height=9cm,angle=0]{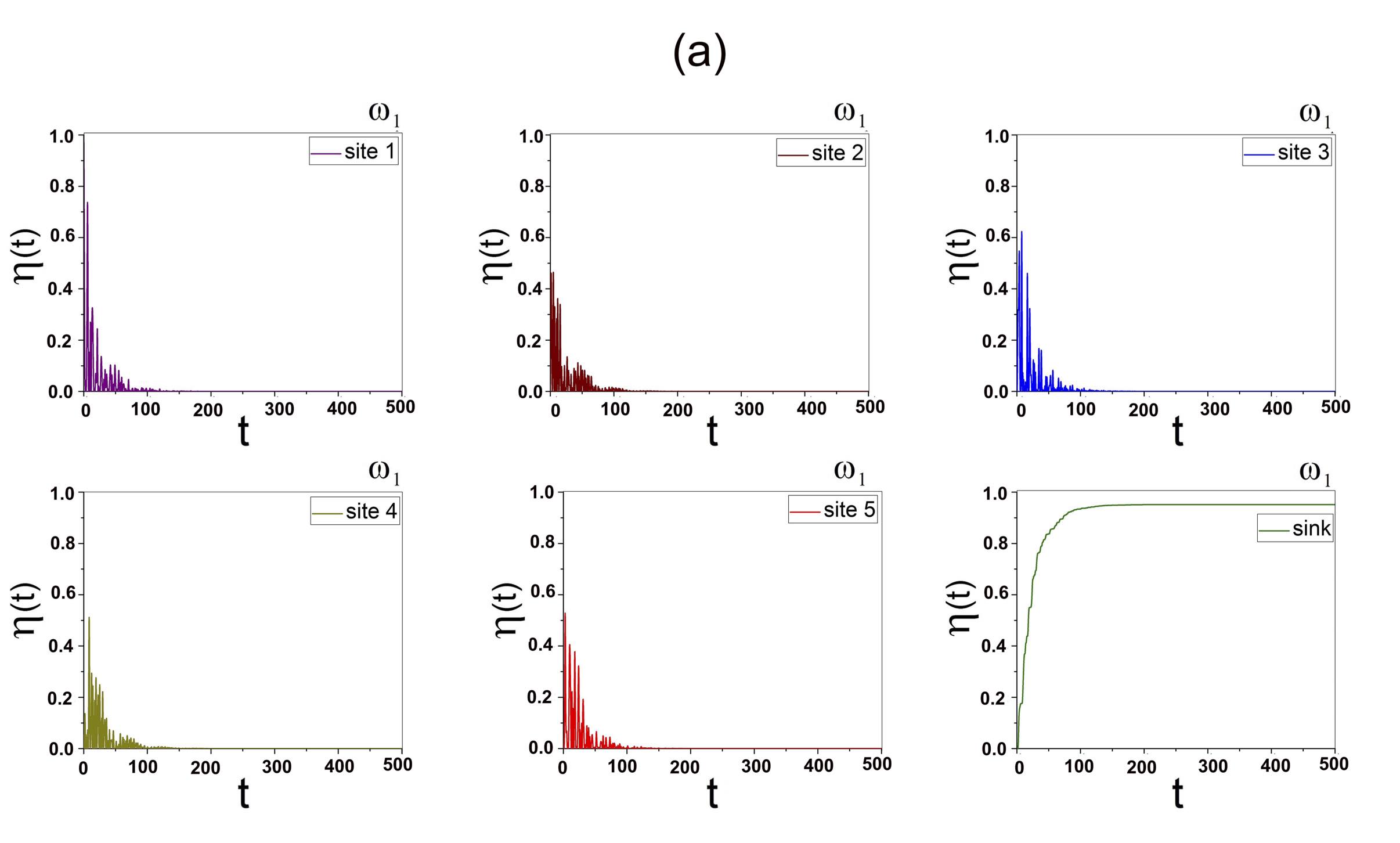}
    \includegraphics[width=16cm,height=9cm,angle=0]{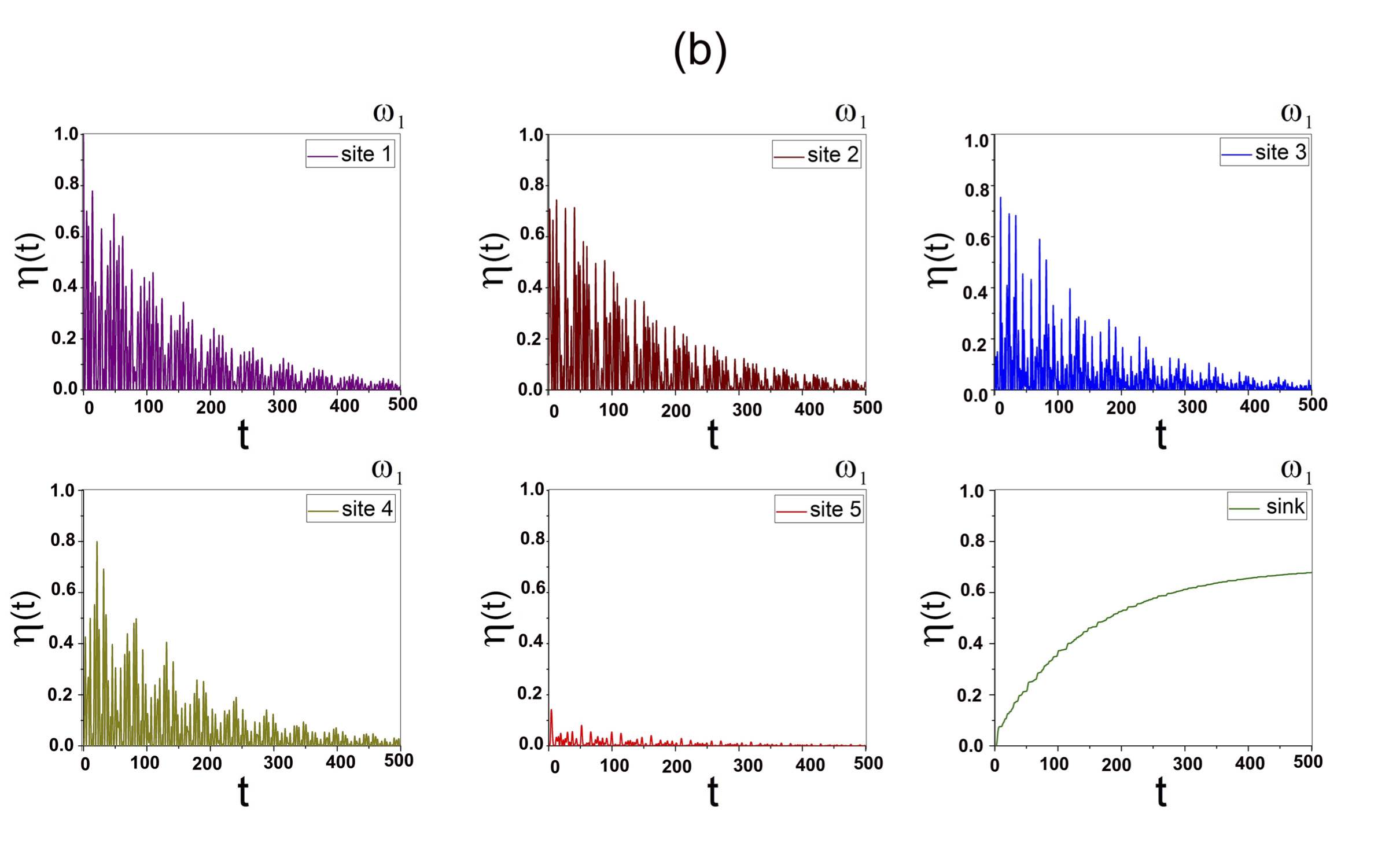}
\caption{ (Colour online) Site populations versus time for $\omega_1$ for (a) an ordered chain and (b) a disordered chain of length $N=5$, for dephasing, 
dissipation, and absorption rates 0, 0.001 and 0.1, respectively.}
\label{fig:fig4}
\end{figure*}
%%%%%%%%%%%%%%%%%%%%%%%%%%% Figure 5 %%%%%%%%%%%%%%%%%%%%%%%%%%%%%%%%%%%%%%%%%%%%%%%%%%%%
 %\begin{figure*}
   %\includegraphics[width=16cm,height=9cm,angle=0]{Fig5a.jpg}
   % \includegraphics[width=16cm,height=9cm,angle=0]{Fig5b.jpg}
%\caption{ (Colour online) Site populations versus time for $\omega_5$ for (a) an ordered chain and (b) a disordered chain of length $N=5$, for dephasing, 
%dissipation, and absorption rates 0, 0.001 and 0.1, respectively.}
%\label{fig:fig5}
%\end{figure*} 
%%%%%%%%%%%%%%% \section{The Role of Noise} %%%%%%%%%%%%%%%

%\subsubsection{The Role of Noise}
\subsection{Effect of the Noise}
\label{noise}

Having discussed the effect of disorder in EET, we now turn our attention to understand how a coupling with the external environment 
may affect the long time behavior of the system. As discussed in the Introduction, biological systems are inherently open and interact with their environment.
This is expected to induce dephasing and affect quantum coherence in the system. 
We will explicitly show below that dephasing (i.e. noise) has no significant effect on population (efficiency) of an ordered chain. However, in the case of a disordered chain
it leads to an enhancement of the EET efficiency.

Noise can be incorporated into the master equation using the following Lindblad form~\cite{plenio.New.J.Phys.2008}:
\begin {equation}
 L_{\rm deph}\rho=\sum_{n=1}^{N}\gamma_{\rm deph}[2{\sigma}^+_n{\sigma}^-_n\rho{{\sigma}^+_n{\sigma}^-_n}-\big\{{{\sigma}^+_n{\sigma}^-_n,\rho}\big\}],
 \label{dephase}
 \end {equation}
where $\gamma_{\rm deph}$ is the dephasing rate coefficient. Thus, the equation of motion for the density matrix becomes
\begin{equation} 
\frac{\partial\rho}{\partial{t}}=i[H,\rho]+L_{\rm diss}\rho+L_{\rm sink}\rho+L_{\rm deph}\rho,
 \label{lin-noise}
 \end{equation}
with the corresponding rates for the three operators.

\begin{figure*}
   \includegraphics[width=13cm,height=8cm,angle=0]{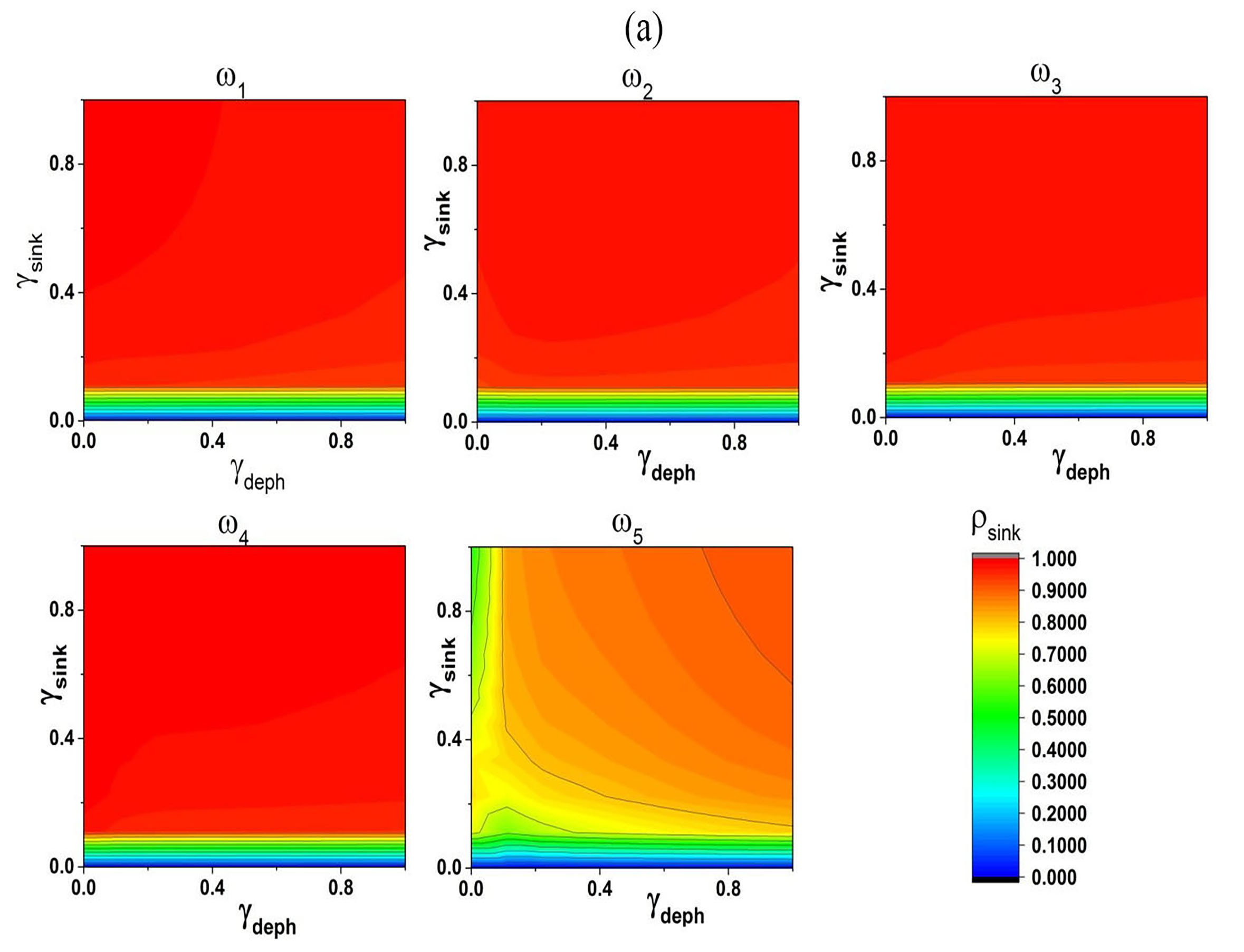}
    \includegraphics[width=13cm,height=8cm,angle=0]{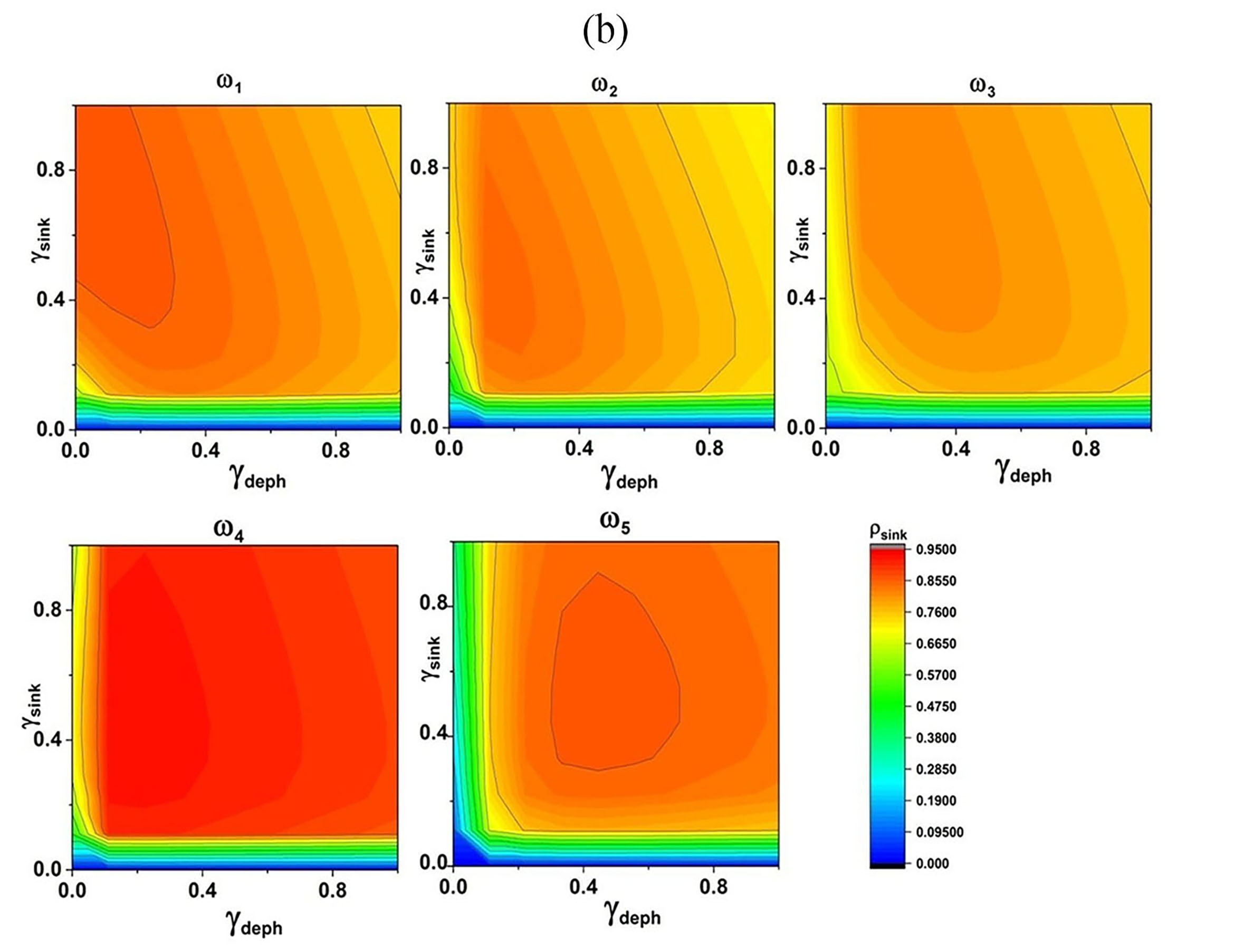}
\caption{ (Color online) Density plots of sink populations $\rho_{\rm sink}$ as a function of the dephasing rate $\gamma_{\rm deph}$ and the absorption 
rate of the sink$\gamma_{\rm sink}$ for 
(a) an ordered chain and (b) a disordered chain with $\gamma_{\rm diss}$= 0.001 for the five different eigenmodes. 
See text for details.} 
  \label{fig:fig5}
\end{figure*}

Figure \ref{fig:fig5} displays contour plots of the steady-state sink populations $\rho_{\rm sink}=\eta(t \to \infty)$ as a function of the dephasing 
$\gamma_{\rm deph}$ and absorption $\gamma_{\rm sink}$ rates. Each plot refers
to a different eigenmodes $\{\omega_1,\ldots,\omega_5\}$ in an ordered (a) and a disordered (b) chain. An interesting pattern emerges from these results.
In general, one would expect a decreasing sink population and/or system efficiency after adding a 
dephasing term to the master equation because of the destructive effect of this term. However, as Fig. \ref{fig:fig5}(a) indicates,  
in a wide range of values of coupling between sink and noise, the value of the sink occupation equals unity 
for the modes from $ \omega_1$ to $ \omega_4$ in an ordered chain.  Thus, noise \textit{improves} EET for these modes.
However, this turns out to be not the case for the mode $\omega_5$ where the site occupation reaches about $0.85$ under optimal conditions. 

Figure \ref{fig:fig5}(b) shows the corresponding results for a disordered system. Similar to an ordered chain, noise enhances EET. The effect is particularly pronounced for
smaller values of $\gamma_{\rm deph}$, and rather pronounced especially for $\omega_4$, where a population in excess of $90\%$ is observed for a wide range of dephasing. 

%%%%%%%%%%%%%%%%%%%%%%%%%%%%%%%%%%% Figure 5 %%%%%%%%%%%%%%%%%%%%%%%%%%%%%%%
%%%%%%%%%%%%%%%%%%%%%%%%%%%%%%%%%%%%%%%%%%%%%%%%%%%%%%%%%%%%%%%%%%%%%%%%%%%%

Additional insights can be obtained by comparing the results in the absence of dephasing in Fig. \ref{fig:fig4} (and Fig.1S) with their counterparts in the presence of noise.
The latter are depicted in Fig. \ref{fig:fig6} (and Fig.2S in the supplementary file) both for an ordered (a) and a disordered (b) chain. In all cases, a value of $\gamma_{\rm deph}=0.2$ for the dephasing rate
was used on the basis of the results reported in Figure \ref{fig:fig5}.
The additional noise here leads to a slightly enhanced efficiency for the
$\omega_4$  eigenmode (not shown), while there is still a relatively large loss for $\omega_5$. An ordered system does not have much backscattering to be 
compensated for by the noise and thus the efficiency is not greatly affected. However, for a disordered system 
depicted in Fig.2S (b) the influence of the noise is much
more pronounced. As a general conclusion it can be stated that the efficiency for all the modes is greatly enhanced here.

The complex behavior of EET observed here for a spatially disordered chain suggests the following physical interpretation that warrants further studies
in the future. In the absence of noise the results clearly show that spatial disorder in the coupling leads to a marked decrease in EET as expected on physical grounds
based on increased scattering. However, the addition of a small or moderate amount of noise quickly recovers the efficiency lost and has a dramatic
effect on improving EET. In our model of five two-level atoms this effect is strongly mode-dependent, too.
The origin for noise-assisted EET in a disordered chain stems from the weakening of localization that reduces the constructive 
coherent backscattering responsible for it \cite{hassan}.

%%%%%%%%%%%%%%%%%%%%% AX jadid %%%%%%%%%%%%%%%%%
%%
%\subsection{Localization with the effect of noise in sites}
%%%%%%%%%%%%%%%%%%%%%%%%%%%%% Figure 6 %%%%%%%%%%%%%%%%%%%%%%%%%%%%%%%%%%
\begin{figure*}
   \includegraphics[width=16cm,height=9cm,angle=0]{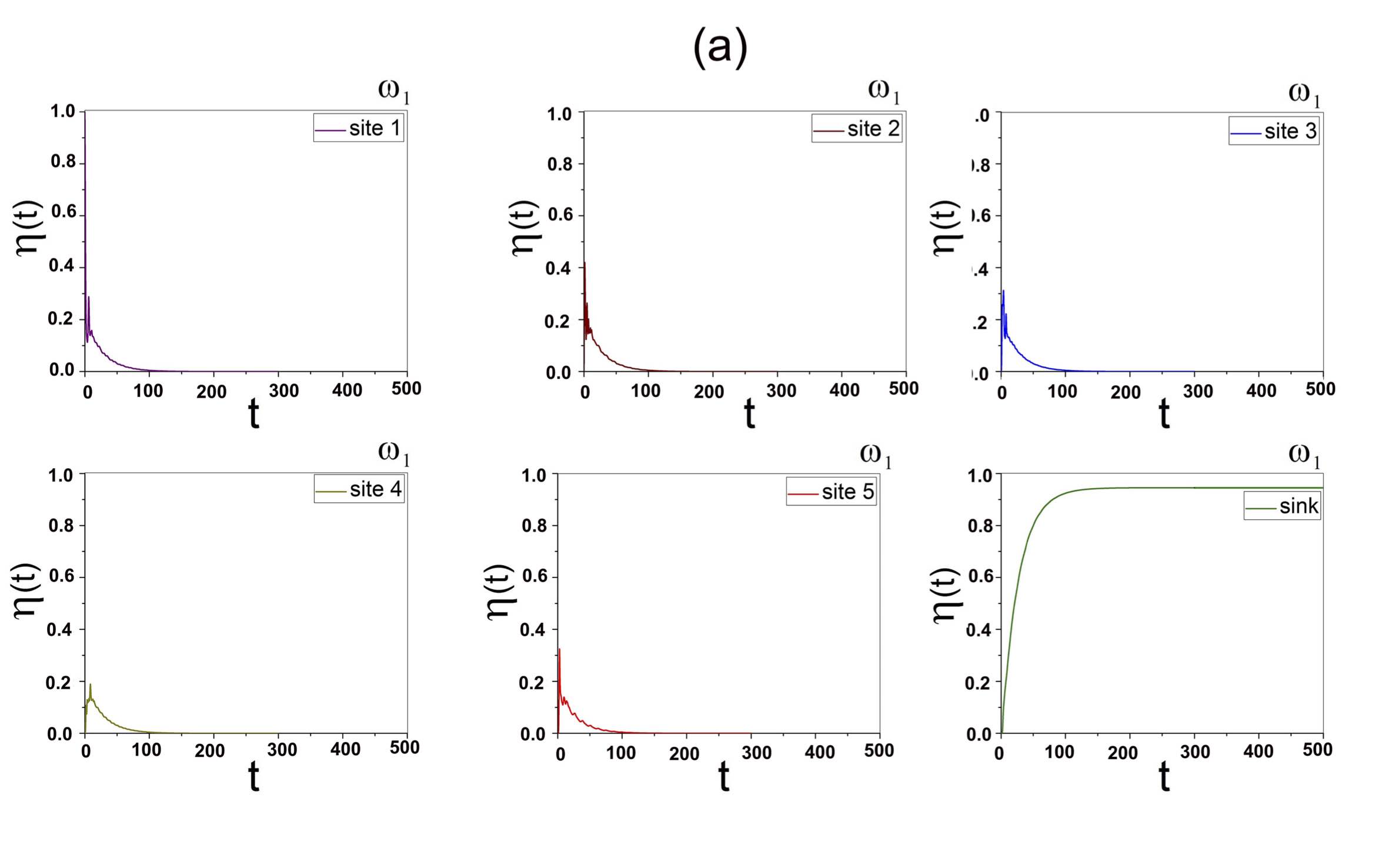}
    \includegraphics[width=16cm,height=9cm,angle=0]{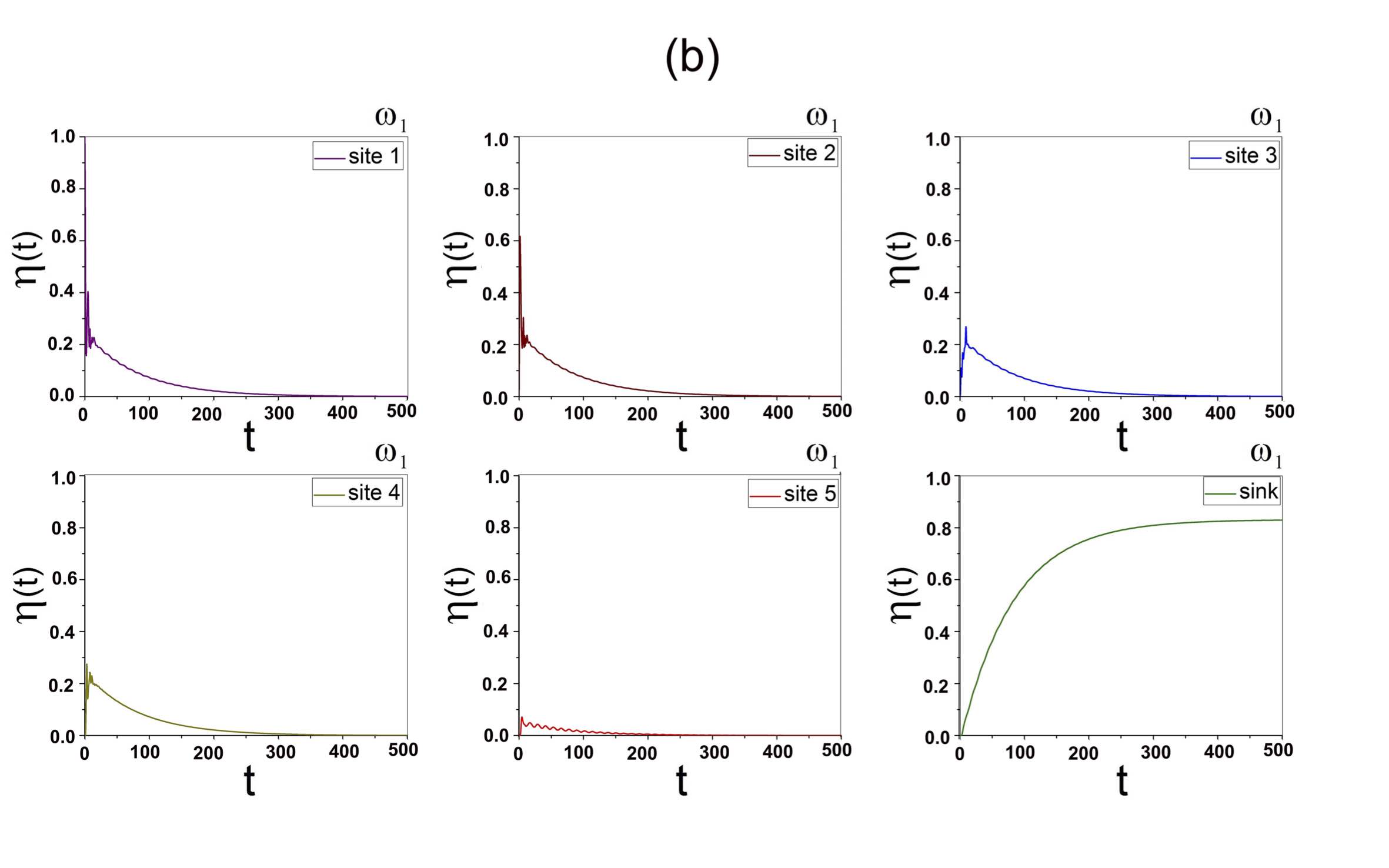}
\caption{ (Colour online) Site populations versus time for $\omega_1$ for (a) an ordered chain and (b) a disordered chain of length $N=5$, for dephasing, 
dissipation, and absorption rates 0.2, 0.001 and 0.1, respectively.}
\label{fig:fig6}
\end{figure*}
%%%%%%%%%%%%%%%%%%%%%%%%%%%%%%%%%%%%%%%%%%%%%%%%%%%%%%%%%%%%%%%%%%%%%
%%%%%%%%%%%%%%%%%%%%%%%%%%%%%% Figure 8 %%%%%%%%%%%%%%%%%%%%%%%%%%%%%%%%%%%%%%
%\begin{figure*}
  % \includegraphics[width=16cm,height=9cm,angle=0]{Fig8a.jpg}
   % \includegraphics[width=16cm,height=9cm,angle=0]{Fig8b.jpg}
%\caption{ (Colour online) Site populations versus time for $\omega_5$ for (a) the ordered chain and (b) the disordered chain of length $N=5$, for dephasing, 
%dissipation, and absorption rates 0.2, 0.001 and 0.1, respectively.}
%\label{fig:fig8}
%\end{figure*} 
%%%%%%%%%%%%%%%%%%%%%%%%%%%%%%%%%%%%%%%%%%%%%%%%%%%%%%%%%%%%%%%%%%%%%%%
%%%%%%%%%%%%%%%%%%%%%%%%%%%% Figure 7 %%%%%%%%%%%%%%%%%%%%%%%%%%%%%%%%
\begin{figure*}
   \includegraphics[width=16cm,height=12cm,angle=0]{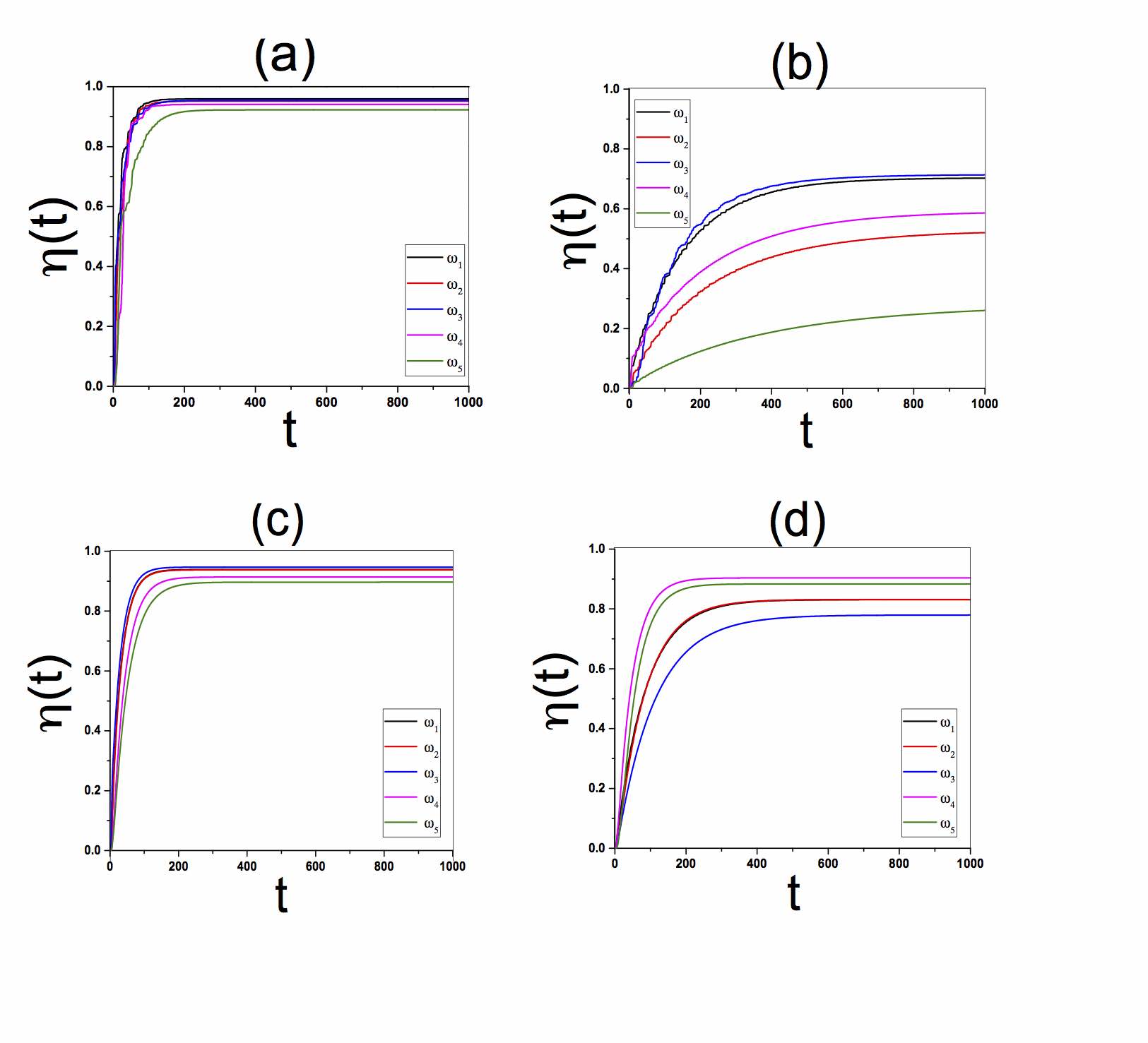}  
\caption{ (Colour online) Sink population versus  time for different normal modes ($\omega_1-\omega_5$) for (a) an ordered chain and (b) a disordered chain of length $N=5$ and dissipation 
(absorption) rate 0.001 (0.1) when there is no dephasing, and (c) an ordered chain and (d) an disordered chain with dephasing rate 0.2 and dissipation (absorption) rate 0.001 (0.1). }
\label{fig:fig7}
\end{figure*}
%%%%%%%%%%%%%%%%%%%%%%%%%%%%%%%%%%%%%%%%%%%%%%%%%%%%%%%%%%%%%%%%%%%%%%%%%%55
%%%%%%%%%%% \section{Summary and Conclusions}%%%%%%%%%%%%
\section{Discussion}

\label{discussion}
%%%%%%%%%%%%%%%%%%%%%%%%%%%%%%%%%%%%%%%%%%%%%%%%%%%%%5
\subsection{Comparison with Real Systems}

Figure \ref{fig:fig7} summarizes our main results by plotting the sink population $\eta(t)$ as a function of time $t$ for both an ordered and a disordered chain, in the absence
(Figs. \ref{fig:fig7} (a) and (b)) and in the presence (Figs. \ref{fig:fig7} (c) and (d)) of noise. Our results suggest that the noise can speed up EET through the strand by about a factor of two 
for a disordered chain,  while it has a significantly less pronounced effect on an ordered chain. This is clearly visible from the values Table II while there is almost no effect on 
EET efficiency for an ordered chain. By contrast, there is a $82.2\%$, $80.4\%$ and $305\%$ EET increment for normal mode frequencies $\omega_2$, $\omega_4$, and $\omega_5$, 
respectively, in a disordered chain.

\begin {table}
\centering

\begin {tabular}{ c  c  c  c  c  c  c }
\hline
Frequency & $\omega_1$ & $\omega_2$ & $\omega_3$& $\omega_4$& $\omega_5$ 
\\ \hline
Ordered chain (no noise)        & 0.94   & 0.91    & 0.95   &0.95  & 0.85
\\ 
 Ordered chain (with noise)            & 0.93  & 0.92   & 0.95    &  0.92 & 0.85
\\ 
Change due to noise ($\%$)         & $-1.1$   & $+1.1$       & 0.0       & $-3.2$  & 0.0
\\ \hline \hline
Disordered chain (no noise)      &    0.70     &   0.45        &   0.73        &  0.51    & 0.22
\\ 
Disordered chain (with noise)         &      0.83    &     0.82      & 0.79 &    0.90      & 0.89
\\ 
Change due to noise  ($\%$)        &   $+18.6$   &  $+82.2$            & $+8.2$  &    $+80.4$    &   $+305$
    
    \end {tabular}
    \caption{The effect of noise on the EET efficiency for the normal mode frequencies of the system. The enhancement of efficiency 
    due to noise in a disorder chain is dramatically higher relative to that in an ordered chain in which noise doesn't play a role.  }
\label{tab11}
\end {table}

A single strand in the selectivity filter backbone is composed of the sequences of TGVYG amino acids (T(Threonine, Thr75), V(Valine, Val76), G(Glycine, Gly77), Y(Tyrosine, Tyr78), G(Glycine, Gly79)) linked by peptide units H–N–C=O. Our model here is a simplified model for above single strand in the selectivity filter, in which the system is made up of 5 sites \cite{hassan}. 
The normal kinetic energy of a single ion in the open state of an ion channel is  $\approx 0.06 {\rm eV} \approx 1.2 {\rm kcal/mol}$ that is two times higher than the thermal energy \cite{10}. Depending on the frequencies of the oscillating C=O pair 
the ion can lose up to half of its kinetic energy which can be transferred to the C=O environment depending on the specific C=O resonant oscillation frequency \cite{Bernroider2}. If we assume that the excitation enrgy in the backbone in the first site is $\approx 0.6 {\rm kcal/mol} $ (i.e. close to the thermal energy) then we estimate the amount of energy at the other end of the strand to be received in less than $1 {\rm ns}$. We have estimated the transferred energy at the sink point for the chain for both noiseless and noisy states (see Table III). It can be seen that the sink energy for noiseless state is in the range $0.132-0.438 {\rm kcal/mol}$ while its value with noise will be $0.474-0.540 {\rm kcal/mol}$. In order to test our model it is enough to measure the energy at the sink point (i.e. last point in the chain, Gly79) in the time interval less than 1ns right after the ion being at the entrance point (i.e. Thr75).  

\begin {table}
\centering

\begin {tabular}{ c  c  c  c  c  c  c }
\hline
Frequency & $\omega_1$ & $\omega_2$ & $\omega_3$& $\omega_4$& $\omega_5$ 
\\ \hline

Sink energy (${\rm kcal/mol}$), no noise      &    0.420     &   0.270        &   0.438        &  0.306    & 0.132
\\ 
Sink energy (${\rm kcal/mol}$), with noise         &      0.498    &     0.492      & 0.474 &    0.540      & 0.534
    
    \end {tabular}
    \caption{The prediction of our model for EET in a single strand in the selectivity filter for different normal frequency modes of the system. It can be seen that the sink energy for noiseless state is in the range $0.132-0.438 {\rm kcal/mol}$ while its value with noise (i.e. more realistic) will be $0.474-0.540 {\rm kcal/mol}$.}
%\label{tab11}
\end {table}

%The dipole-dipole coupling between two peptide units in a P-loop monomer can be around $\approx 0.15 {\rm eV} \approx 1210 {\rm Cm^{-1}}$ \cite{}. When an ion gets very close to a carbonyl group, it forces the C=O to align its dipole with its electric field. Moreover, the dipole-dipole interactions between adjacent carbonyls cause a longitudinal oscillation through the monomer chain. The energy of coupling between a potassium ion and a carbonyl group is $E_{\rm C=O, K}=2-3 eV \approx 50\times 10^{-20}$ J \cite{Bucher} and the energy of a C=O vibration is $E_{\rm C=O, v}\approx 4\times 10^{-20}$ J. 

Molecular Dynamics simulations have shown that the C=O bond in a carbonyl group may oscillate radially and angularly with typical amplitudes of 
$a^{{\rm C=O}, r}_n=0.005$ {\AA} and $a^{{\rm C=O}, \theta}_n=0.01$ {\AA}, respectively, and they have electrical dipoles with dipole moments 
$P_{\rm C=O}=7.2\times 10^{-30}$ Cm \cite{Vahid2} where the distance between the effective charges of the C=O dipoles is approximately $0.122$ nm\cite{Bernroider2}. 
Basically, biomolecular complexes are flexible structures subjected to significant temperature fluctuations with amplitudes on the order of $0.75$ to $1.0$ {\AA} \cite{Gwan}. 
The resonance frequency of carbonyl is approximately $\omega_{\rm C=O} \approx 10^{14}$ Hz, corresponding to a wave number of about $1700$ cm$^{-1}$ 
where the oscillation is essentially in the ground state at room temperature \cite{Vahid2, plenio.New Journal of Physics2010}. 
The natural oscillation frequency of the C=O groups caused by thermal energy is essentially determined by the binding strength of the C=O groups to the 
backbone \cite{Bernroider2}. 
The concerted quantum-mechanical behavior of the ion together with the coordinating carbonyl dipoles can account for the unique selection procedure for ion selectivity \cite{Bernroider2}.

If the system-environment interaction leads to dissipation the strength of the system-environment interaction is a measure of the relaxation time. 
As the system-environment interaction strength decreases, the relaxation times become longer, and vice versa.\cite{iiiii} The previous estimates 
in the selectivity filter indicate that the relaxation times are mainly of the order of nanoseconds but the decoherence times are in the picoseconds range \cite{iiiii}.

 \subsection{Limitations of the Model}
 
 Despite many attempts the details of the operation mechanisms underlying ion selectivity and transport in the ion channel are still not fully understood, 
probably due to the lack of experimental methods to probe the mechanisms dynamically on the biologically relevant times cales \cite{Vaz2}. 
The dynamics inside the selectivity filter is indeed a many-body phenomenon involving many interacting degrees of freedom, and can hardly be reduced to the 
simple model Hamiltonians used in the present study at a quantitative level. For example, the expression (4) of the sink population assumes that all excitations in the last site succeed in leaving the system \cite{plenio.New Journal of Physics2010}. Hence, only qualitative conclusions can be drawn with the present approach. Despite these shortcomings
we expect that the main result of this study - how the noise can enhance EET in the selectivity filter - should hold even when more quantitative models are considered.
At present we are unaware of experimental evidence for non-trivial quantum effects in ion channels, except for indirect evidence based on femtosecond laser 
spectroscopy techniques such as 2D-IR, FT-IR, etc. \cite{Vaz1, Vaz2}. We further note that in real systems the distances $d_n (t)$ are expected to be randomly fluctuating 
functions of time and not harmonic functions, and the issue of whether these thermal fluctuations are comparable with the amplitudes of the eigenmodes is an open issue.
 
%%%%%%%%%%%%%%%%%%%%%%%%%%%%%%%%%%%%%%% 
 \section{Conclusions and Future Perspectives}
 \label{conclusions}
%%%%%%%%%%%%%%%%%%%%%%%%%%%%%%%%%%%%%%
 Because of the small size ($0.3$ nm in width and $1.2$ nm in length), small time scales (ps and ns) and small energy scales of the transport processes ($0.5-3$ eV \cite{Bucher}) a 
 selectivity filter can be considered as a nanoscale protein machine \cite{Gadsby} in which quantum coherence effects might play a functional role \cite{Vaz2}. It has 
 been suggested that quantum coherence and its interplay with thermal vibrations might be involved in mediating ion selectivity and transport \cite{Vaz2}. 
 The translocation of ions between the sites occurs at the picosecond time scale which implies that quantum coherence may appear in the system 
 and the mechanism behind resonances in the filter may not follow classical rate equation dynamics \cite{plenio.New Journal of Physics2010}. 
 It has been shown that classical coherence among carbonyl vibrations of the backbone in the selectivity filter plays no 
 significant role in fast ionic conduction \cite{Vahid2}. 
%\label{conclusion}
In this work, we have considered harmonic chains of oscillating particles as simple models for a P-loop strand of the 
selectivity filter backbone in biological ion channels. The EET in such systems was modeled by a tight-binding 
Hamiltonian in the weak coupling Markovian master equation framework. We considered both the case of a fully
symmetric chain of $N=5$ two-level atoms coupled to a sink at the end of the chain, and the case where the distance between the nearest
neighbor atoms is not identical (the so-called disordered chain). A dipole-dipole interaction was assumed between nearest
neighbors in which the couplings between dipoles are changing in time due to intrinsic dynamics. Our results for EET in these chains for the various 
eigenmodes reveal that while spatial disorder induces
localization and suppresses EET in the disordered chain, adding noise has a constructive influence for a relatively wide range
of system parameters. This supports the view that noise which is inherent in real systems can greatly facilitate
EET in a biological selectivity filter which is a disordered system but very efficient. In this sense, dephasing may in fact assist the transport 
as it suppresses destructive interference at the resonance points, which is a non-classical effect. Thus noise might have an important role for efficient energy transfer
through their structures for a better coordination of ion movements \cite{Semiao, Gustav, Bernroider2, Vaz1, Vaz2, Vahid2} and transfer and filter gating \cite{Bernroider2}.

%\section{Future prospects}
The model presented here allows for several possible improvements. First, an extension to include all four strands of the selectivity filter would allow to 
capture the effects of the topology of the system. Support from dedicated experiments would make this task easier.
To this aim, we suggest to combine terahertz techniques with a highly sensitive X-ray crystallographic method to visualize low-frequency vibrational modes in the structure \cite{Lundholm}. 
Regarding disorder effects, 2D IR spectroscopy can estimate the amount of structural disorder by observing the diagonal elongation in the spectral peaks \cite{Vaz2}. 
Second, it would be interesting to include the effect of the solvent (water). We hope to be able to address both issues in future works.
%\end{widetext}

\appendix
\section{Supplementary Figures}
%%%%%%%%%%%%%%%%%%%%%%%%%%% Figure S1  %%%%%%%%%%%%%%%%%%%%%%%%%%%%%%%%%%%%%%%%%%%%%%%%%%%%
\begin{figure*}
  \includegraphics[width=16cm,height=9cm,angle=0]{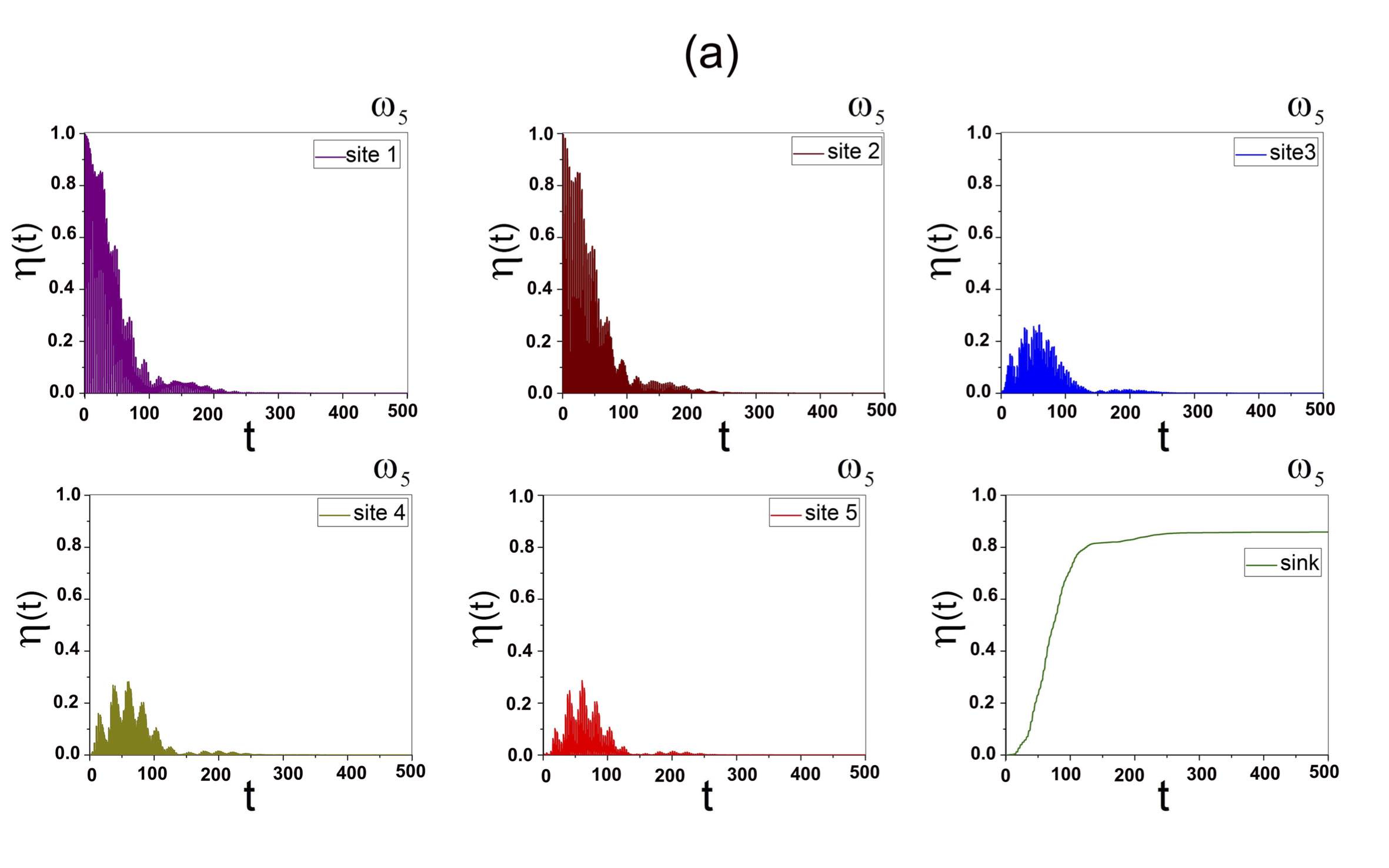}
    \includegraphics[width=16cm,height=9cm,angle=0]{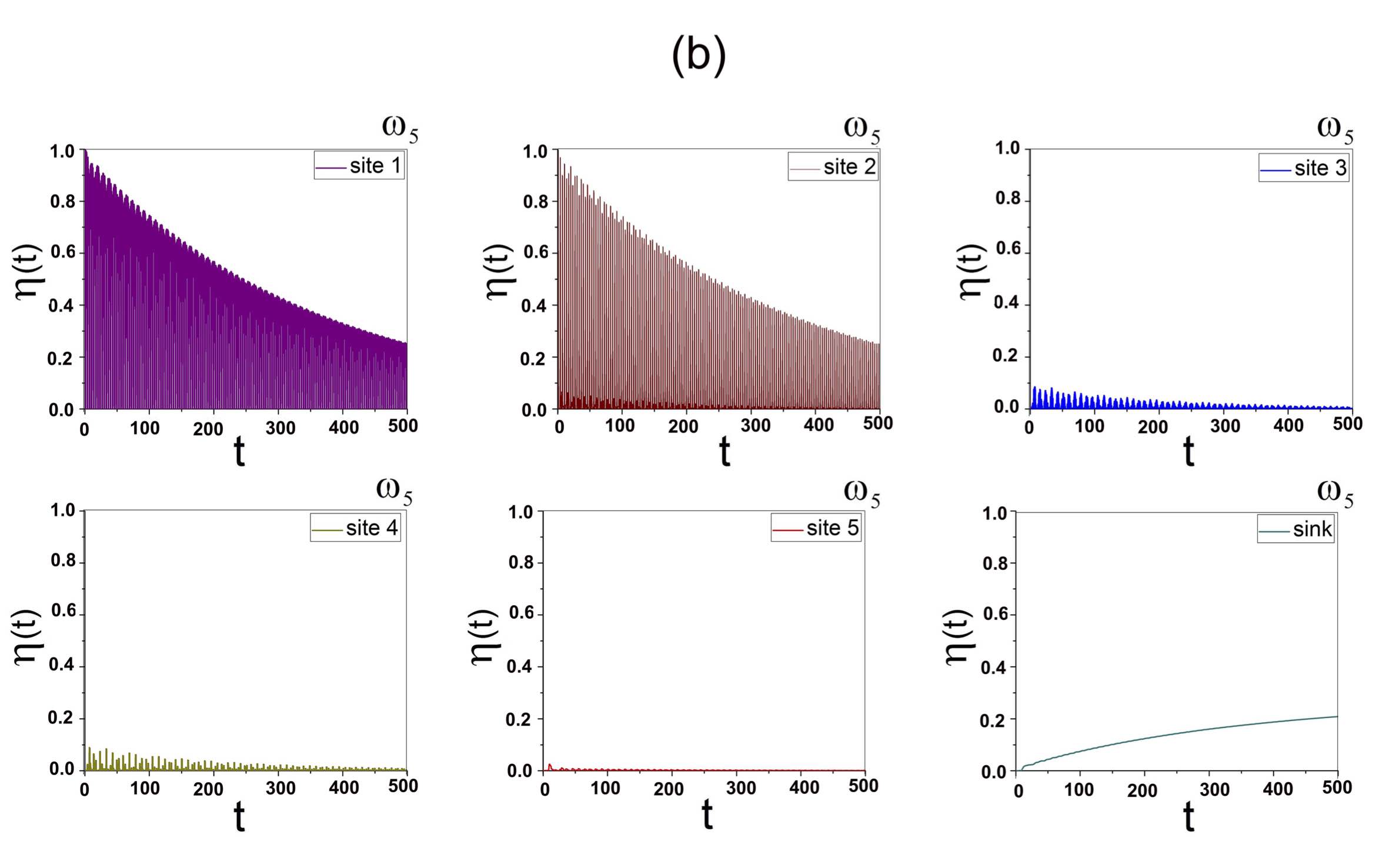}
\caption{ (Colour online) Site populations versus time for $\omega_5$ for (a) an ordered chain and (b) a disordered chain of length $N=5$, for dephasing, 
dissipation, and absorption rates 0, 0.001 and 0.1, respectively.}
\label{fig:figS1}
\end{figure*} 
%%%%%%%%%%%%%%% %%%%%%%%%%%%%%%

%%%%%%%%%%%%%%%%%%%%%%%%%%%%%% Figure S2 %%%%%%%%%%%%%%%%%%%%%%%%%%%%%%%%%%%%%%
\begin{figure*}
  \includegraphics[width=16cm,height=9cm,angle=0]{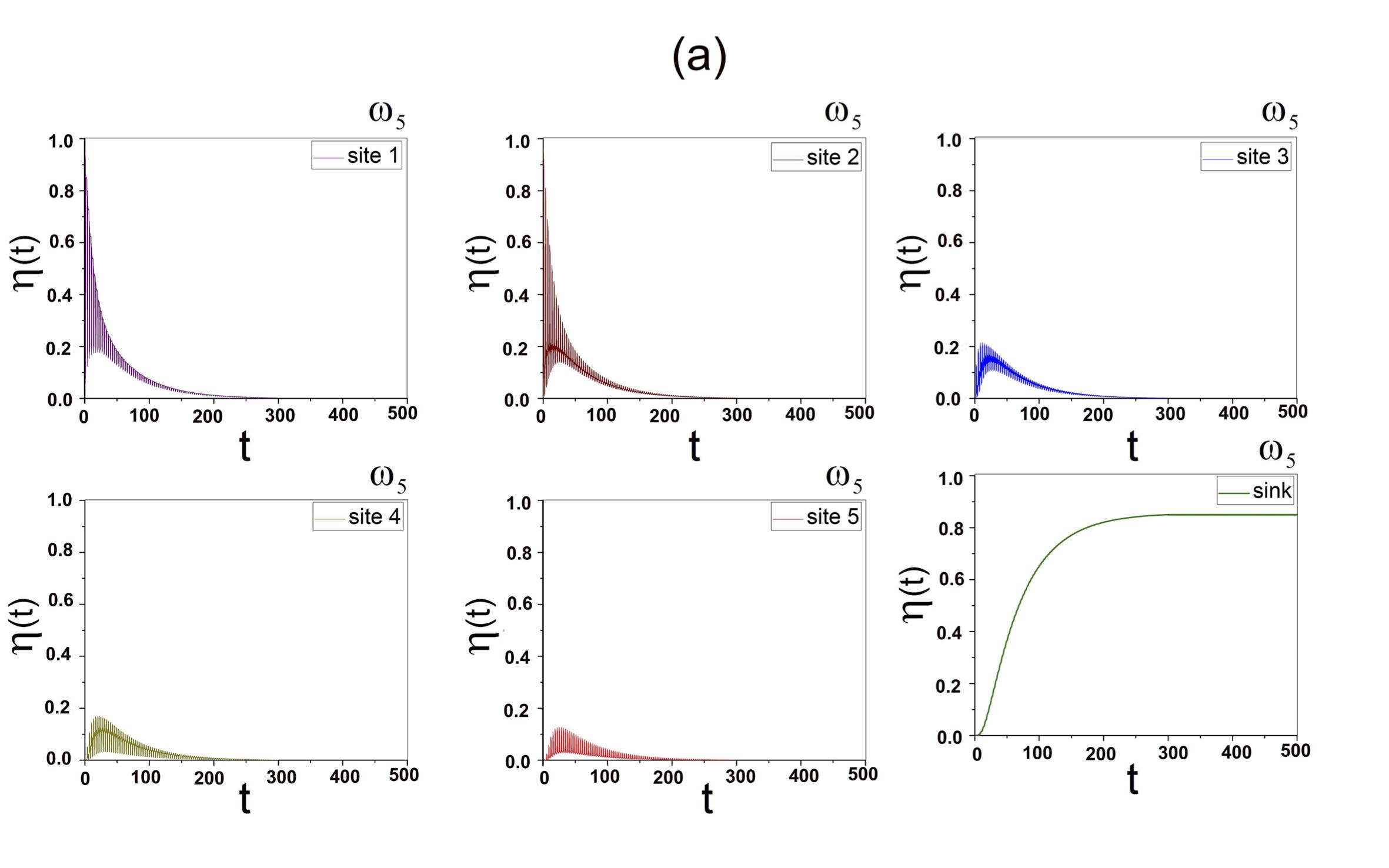}
   \includegraphics[width=16cm,height=9cm,angle=0]{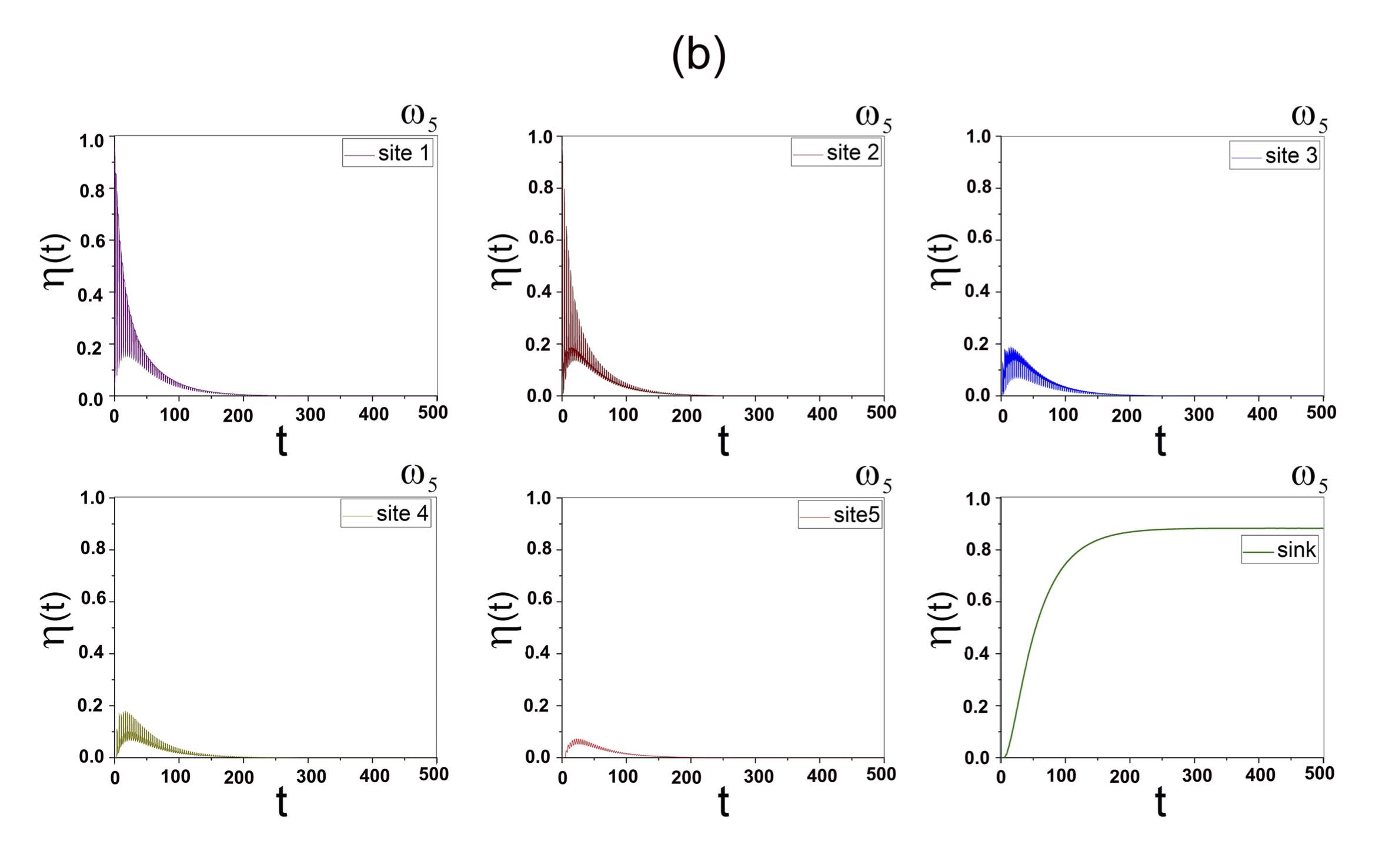}
\caption{ (Colour online) Site populations versus time for $\omega_5$ for (a) the ordered chain and (b) the disordered chain of length $N=5$, for dephasing, 
dissipation, and absorption rates 0.2, 0.001 and 0.1, respectively.}
\label{fig:figS2}
\end{figure*} 
%%%%%%%%%%%%%%%%%%%%%%%%%%%%%%%%%%%%%%%%%%%%%%%%%%%%%%%%%%%%%%%%%%%%%%%

\clearpage

%%%%%%%%%%%% References and Notes %%%%%%%%%%%%%%%%%%%%


\begin{thebibliography}{60}
\bibitem{yyyy} Here we set the total length of the both the ordered and the disordered chains equal. 
Thus the sum $\sum d_{n}$ for both ordered and disordered should be equal and we find $d_0=1.21$ for the ordered chain.
\bibitem{footnote1}{Note that it is possible that
each two-level system (atom) feels a different environment and hence the dissipation rates $\gamma$ could depend on $n$. 
However, it has been shown in Ref. \onlinecite{hassan} that role of dissipation is exponentially destructive for such chains and thus
we assume here $\gamma$ to be identical on all sites.}
\bibitem{footnote2}{Strictly speaking the hopping integrals could also be time dependent but here we neglect this complication for simplicity.}
\bibitem{python}{In all the calculations in this work, we used QUTIP\cite{qutip} to numeri- cally solve the master equations and Sympy\cite{sympy} to obtain the normal modes.}
\bibitem{footnote3}{Strictly speaking the normal modes should be recalculated numerically for the disordered chain,
but this would not change our conclusions here.}
\end{thebibliography}
\end {document}